\documentclass[preprint,aps,showpacs,showkeys,amsmath,amssymb,nopreprintnumbers,nofootinbib]{revtex4}
\usepackage{graphicx}% Include figure files
\usepackage{dcolumn}% Align table columns on decimal point
\usepackage{bm}% bold math
\usepackage[dvipdfm]{color}
\usepackage[dvipdfm,bookmarks=true,bookmarksnumbered=true,bookmarkstype=toc]{hyperref}

\def\Rnum#1{\resizebox{0.5em}{\height}{\uppercase\expandafter{\romannumeral #1}}}

\begin{document}

%\preprint{}

\title{Properties of dense, asymmetric nuclear matter in Dirac-Brueckner-Hartree-Fock approach}%

\author{Tetsuya Katayama}
\email{j6213701@ed.tus.ac.jp}
\affiliation{%
Department of Physics, Faculty of Science and Technology,\\
Tokyo University of Science, Noda 278-8510, Japan 
}%
\author{Koichi Saito}
\email{koichi.saito@rs.tus.ac.jp}
\altaffiliation[Also at ]{J-PARC Branch, KEK Theory Center, Institute of Particle and Nuclear Studies, KEK, Tokai 319-1106, Japan}
\affiliation{%
Department of Physics, Faculty of Science and Technology,\\
Tokyo University of Science, Noda 278-8510, Japan 
}%

\date{\today}% It is always \today, today,
             % but any date may be explicitly specified

\begin{abstract}
Within the Dirac-Brueckner-Hartree-Fock approach, using the Bonn potentials, we investigate the properties of dense, asymmetric nuclear matter 
and apply it to neutron stars.  
In the actual calculations of the nucleon self-energies and the energy density of matter, we study in detail the validity of an angle-averaged approximation and an 
averaging of the total momentum squared of interacting two-nucleons in nuclear matter.  For practical use, we provide convenient parametrizations for the equation of state 
for symmetric nuclear matter and pure neutron matter.  We also parametrize the nucleon self-energies in terms of polynomials of nucleon momenta.  
Those parametrizations can accurately reproduce the numerical results up to high densities. 
\end{abstract}


%Use showkeys class option if keyword
\maketitle
%%%%%%%%%%%%%%%%%%%%%%%%%%%%%%%%%%%%%%%%%%%%%%%%%%

\section{Introduction}

Neutron stars are the most dense and exotic states of nuclear matter in the universe, and it is generally believed that the central density in massive neutron stars 
reaches several times higher than the normal nuclear density, $n_B^0$.  
Consequently, not only nucleons and leptons but also hyperons may exist stably in the core of neutron star.  
Because of this extreme situation, observations of neutron stars can provide very important information which cannot be obtained from terrestrial experiments, 
and thus have recently attracted much attention.  

However, the detail of the properties of the core is still in the middle of an argument, although there have been extensive efforts to study it 
experimentally and theoretically.  Because the appearance of hyperons in dense nuclear matter inevitably softens the equation of state (EoS), it is believed that 
the maximum neutron-star mass becomes seriously diminished if hyperons are created in the core of neutron star. 
As a result, the majority of theoretical approaches including hyperons, in which Quantum hadrodynamics (QHD) \cite{Serot} in Hartree approximation 
(or mean field theory (MFT)) is mostly used, is ruled out by the reliable determination of the mass of pulsar J1614-2230, namely 
$1.97\pm0.04M_{\odot}$ ($M_{\odot}$: the solar mass) \cite{Demorest}.  
(Recently, a new massive pulsar, J0348+0432, has been reported, and the mass is estimated to be $2.01 \pm 0.04 M_{\odot}$ \cite{Antoniadis}.) 

However, before concluding that the EoS including hyperons cannot explain the heavy neutron stars, it is very important to study the effect of in-medium 
two-nucleon correlations, which depends on the nuclear density, $n_B$,  
on the EoS for the core matter.  In this paper, we calculate the properties of dense, asymmetric nuclear matter using 
the Dirac-Brueckner-Hartree-Fock (DBHF) approach \cite{Horowitz,Haar,Brockmann,Sehn,Gross,Dalen,Klahan,Krastev}, 
which allows us to handle two vital ingredients: short-range correlations in nucleon-nucleon (NN) force, which are not included in MFT, and relativistic many-body effects.

In MFT, in spite of the considerable uncertainty of nuclear matter properties, the meson-baryon coupling constants are usually adjusted so as to reproduce such 
data on nuclear matter around $n_B^0$.  Then, the same coupling constants (without density dependence) are mostly used even in the calculation for neutron stars\footnote{
There are some exceptions. For example, 
in the density-dependent meson-exchange (DDME) model \cite{DDME}, the coupling constants depend on $n_B$, and those 
are parametrized in terms of some specific functions.  In the quark-meson coupling (QMC) model \cite{katayama}, the $\sigma$-baryon couplings depend 
on the scalar density in matter.}.  
Therefore, such extrapolation to extremely isospin-asymmetric and dense matter like neutron stars may be considered with skepticism.

In contrast, the DBHF calculation is based on the realistic NN interactions, which have been determined using a large amount of NN scattering data, and it can mostly reproduce 
the nuclear matter properties around $n_B^0$ without any readjustment of the coupling constants\footnote{ 
We note that the values of coupling constants used in MFT are, of course, quite different from those in the DBHF calculation.}. 
The reason for the great success of this method is the density dependence of the in-medium NN scattering amplitude mainly caused by the Pauli exclusion principle, 
the short-range NN correlations and many-body forces.  Thus, it must be important even in the calculation for neutron stars. 

However, the DBHF approach is quite complicated and demands enormous time for the calculation.  As a result, it is not easy to perform the DBHF calculation for neutron stars 
even without hyperons, much less the inclusion of hyperons is quite difficult.  
Thus, it is very vital to reduce the calculation time and costs in order to actually perform the full DBHF calculation including hyperons in the future.  

For such purposes, in this work, we study a dimension-reduction approximation in multi-dimensional integrals.  In the DBHF approach, it is necessary to 
work out the coupled, two-dimensional integral equations to obtain the nucleon self-energies.  Furthermore, to calculate the energy density of matter, we have to solve  
the three-dimensional integrals. Such calculations actually require enormous time.  Then, to reduce the dimension of integrals and carried out the calculation with less 
computation time, we exploit an angle-averaged approximation and an averaging of the total momentum squared of interacting two-nucleons in nuclear matter. 
In particular, we find that the angle-averaged approximation, which has for the first time been used in the DBHF calculation by Horowitz and Serot \cite{Serot,Horowitz} 
to study the nuclear matter properties around $n_B^0$, is very useful and accurate even in the calculation for dense nuclear matter.  
We then calculate the mass-radius relation for neutron stars solving the Tolman-Oppenheimer-Volkoff (TOV) 
equation \cite{TOV}, and examine the validity of those approximations in the calculation for neutron stars. 
 
For practical use, we also provide convenient parametrizations for the EoSs 
for symmetric nuclear matter and pure neutron matter.  In addition, we parametrize the nucleon self-energies in terms of polynomials of nucleon momenta.  
Those parametrizations can accurately reproduce the numerical results up to high densities ($\sim 1.2$ fm$^{-3}$). 

The paper is organized as follows.
In Sec. \ref{sec:DBHF_approach}, the DBHF approach is reviewed briefly.  In Secs. \ref{sec:nuclear_matter_properties} , \ref{sec:dimension_reduction} and \ref{sec:parametrization},  
we respectively present the numerical results for the properties of symmetric nuclear matter, the validity of the dimension-reduction approximation  
and the parametrizations for the nucleon self-energies and the EoSs for nuclear matter.  
Finally, we give a summary and conclusions in Sec. \ref{sec:summary_and_conclusion}.  
Throughout this work, we shall use the notation conventions used in Ref. \cite{Horowitz}. 

%%%%%%%%%%%%%%%%%%%%%%%%%%%%%%%%%%%%%%%%%%%%%%%%%%
\section{Dirac-Brueckner-Hartree-Fock calculation\label{sec:DBHF_approach}}

In the DBHF approximation, the nucleon in asymmetric nuclear matter is described by a set of three, simultaneous integral equations: 
the Bethe-Salpeter equations, single-particle proper self-energies and Dyson's equations. 
The two-nucleon correlations in matter are then expressed by the effective interactions, $\Gamma$, which are given by the solutions to the coupled, 
ladder-approximated Bethe-Salpeter equations: 
\begin{eqnarray}
	\Gamma_{ii}^{dir,ex}&=&V_{ii}^{dir,ex}+\int V_{ii}^{dir,ex}Q_{ii}^{av}\tilde{g}_{ii}\Gamma_{ii}^{dir,ex} \ ,\label{eq:Gammaii}\\
	\Gamma_{ij}^{dir}&=&V_{ij}^{dir}+\int V_{ij}^{dir}Q_{ij}^{av}\tilde{g}_{ij}\Gamma_{ij}^{dir}+\int V^{ex}_{ij}Q_{ji}^{av}\tilde{g}_{ji}\Gamma_{ij}^{ex} \ ,\\
	\Gamma_{ij}^{ex}&=&V_{ij}^{ex}+\int V_{ij}^{ex}Q_{ij}^{av}\tilde{g}_{ij}\Gamma_{ij}^{dir}+\int V^{dir}_{ij}Q_{ji}^{av}\tilde{g}_{ji}\Gamma_{ij}^{ex} \ , \label{eq:Gammaijex}
\end{eqnarray}
where $i$ or $j$ denotes neutron ($n$) or proton ($p$), and the superscript, $dir \, (ex)$, stands for the direct (exchange) term in the two-nucleon scattering.  
The ladder-approximated Bethe-Salpeter equations %, which represent the short-range NN correlations, 
sum to all orders 
the one-boson-exchange (OBE) interactions, $V$, between two nucleons. Here, $Q^{av}$ is the angle-averaged Pauli exclusion operator, 
and $\tilde{g}$ denotes an approximated two-nucleon propagator \cite{Horowitz}.  The use of these quantities allows us to reduce 
the four-dimensional integral equations, Eqs.~(\ref{eq:Gammaii}) - (\ref{eq:Gammaijex}), to two-dimensional ones, 
which makes the further numerical calculation in the center-of-mass frame of the interacting two nucleons feasible.  
In the present calculation, as $\tilde{g}$, we use the Thompson propagator, $\tilde{g}_{Th}$ \cite{Thompson}.  

The self-energy, $\Sigma$, is connected to the effective interaction in a nuclear medium by summing up all orders of direct and exchange diagrams. It is given by
\begin{equation}
	\Sigma_i=-i\sum_{j=p,n}\int\left[\mathrm{tr}(G_j\Gamma_{ij}^{dir})-G_j\Gamma_{ij}^{ex}\right] \ , \label{eq:Self_energy}
\end{equation}
where $G_j$ is the Green's function of interacting nucleon.
In addition to Eqs. (\ref{eq:Gammaii})-(\ref{eq:Self_energy}), to impose self-consistency, the Dyson's equation must be satisfied: 
\begin{equation}
	G_i = G^0_i + G^0_i \Sigma_i G_i \ ,
\end{equation}
where $G_i^0$ is the Green's function in free space.
If the higher-order diagrams in Eqs. (\ref{eq:Gammaii})-(\ref{eq:Gammaijex}) are ignored, the above procedure coincides with relativistic Hartree-Fock approximation. 
In addition, if we ignore the exchange diagram in Eq. (\ref{eq:Self_energy}), the relativistic Hartree (or MFT) result is reproduced.  

The self-energy in a uniform matter can generally be written as
\begin{equation}
	\Sigma_i(k)=\Sigma^S_i(k)-\gamma^0\Sigma^0_i(k)+{\boldsymbol\gamma}\cdot{\bm k}\Sigma^V_i(k),
\end{equation}
with ${\bm k}$ ($k$) being the three- (four-) momentum.  Here, $\Sigma^{S (0) [V]}$ is the scalar (the zero-th component of vector) [the space component of vector] part of the 
nucleon self-energy.  
The actual calculation for these self-energy components is explained in Appendix \ref{sec:Self_energy_S0V}.  

Using the self-energies,  %By dividing the self-energy into three terms: $\Sigma^{S(0)[V]}_i$, 
the effective nucleon mass, $M_i^{\ast}$, and the single-particle energy, $E_i(k)$, in matter are obtained as 
\begin{eqnarray}
	M_i^{\ast}&=&\frac{M_N+\Sigma_i^S(k_{Fi},E_i^{\ast}(k_{Fi}))}{1+\Sigma_i^V(k_{Fi},E_i^{\ast}(k_{Fi}))} \ ,\\
	E_i(k)&=&E^{\ast}_i(k)-\Sigma^0_i(k_{Fi},E_i^{\ast}(k_{Fi}))+E^{\ast}_i(k)\Sigma_i^V(k_{Fi},E_i^{\ast}(k_{Fi})) \ ,
\end{eqnarray}
with $M_N$ being the free nucleon mass, $E_i^{\ast}(k)=\sqrt{{\bm k}^2+M_i^{\ast}}$ and $k_{Fi}$ being the Fermi momentum of nucleon $i$.  
The Dirac equation in matter then reads 
\begin{equation}
	\left[{\boldsymbol \alpha}\cdot{\bm k}+\beta M_i^{\ast}\right]u_i({\bm k},\lambda)=E_i^{\ast}(k)u_i({\bm k},\lambda) \ ,
\end{equation}
with the solution with helicity, $\lambda$, 
\begin{equation}
	u_i({\bm k},\lambda)=\sqrt{E^{\ast}_i(k)+M_i^{\ast}}\left(
	\begin{array}{c}
	1\\
	\frac{2\lambda \underline{k}}{E_i^{\ast}(k)+M_i^{\ast}}
	\end{array}
	\right)\chi_{\lambda} \ , \label{eq:spinor}
\end{equation}
where $\underline{k}\equiv|{\bm k}|$ and $\chi_{\lambda}$ is the Pauli spinor. This solution is normalized as $\bar{u}_iu_i=2M_i^{\ast}$.

The energy density of nuclear matter, ${\cal E}$, is composed of the kinetic and potential energies, which are given in terms of $\Gamma$ as
\begin{equation}
	{\cal E}=\sum_{i=p,n}\left(\left<\hat{\cal T}_{i}\right>+\sum_{j=p,n}\left<\hat{\cal V}_{ij}\right>\right) \ ,
\end{equation}
with
\begin{eqnarray}
	\left<\hat{\cal T}_{i}\right>& = &\frac{k_{Fi}}{8\pi^2}(k_{Fi}^2+M_i^{\ast2})^{1/2}[4M_iM_i^{\ast}-3M_i^{\ast2}+2k_{Fi}^2]\nonumber\\
	&&+\frac{M_i^{\ast3}}{8\pi^2} [3M_i^{\ast}-4M_i] \log\left(\frac{k_{Fi}+(k_{Fi}^2+M_i^{\ast2})^{1/2}}{M_i^{\ast}}\right) \ , \\
	\left<\hat{\cal V}_{ij}\right>& = &\int\frac{d^3p}{(2\pi)^3}\int\frac{d^3P}{(2\pi)^3}\frac{\theta(k_{Fi}-\left|\frac{1}{2}{\bm P} 
 +{\bm p}\right|)}{2E_i^{\ast}(\frac{1}{2}{\bm P}+{\bm p})}\frac{\theta(k_{Fj}-\left|\frac{1}{2}{\bm P}-{\bm p}\right|)}{2E_j^{\ast} 
(\frac{1}{2}{\bm P}-{\bm p})}\tilde{\Gamma}_{ij} \ , \label{eq:Vij}
\end{eqnarray}
where ${\bm P}$ and ${\bm p}$ are, respectively, the total and relative momenta for the interacting two nucleons in the matter-rest frame.  
Using the spinor, Eq. (\ref{eq:spinor}), in that frame, the matrix element, $\tilde{\Gamma}_{ij}$, is calculated by 
\begin{eqnarray}
	\tilde{\Gamma}_{ij}& = &\sum_{\lambda\lambda^{\prime}}\left[\bar{u}_i\left(\frac{1}{2}{\bm P}+{\bm p},\lambda\right) 
 \bar{u}_j\left(\frac{1}{2}{\bm P}-{\bm p},\lambda^{\prime}\right) \Gamma_{ij}^{dir} u_i\left(\frac{1}{2}{\bm P}+{\bm p},\lambda\right)u_j\left(\frac{1}{2}{\bm P}-{\bm p},\lambda^{\prime}\right)\right. \nonumber\\
	&&\left.-\bar{u}_i\left(\frac{1}{2}{\bm P}+{\bm p},\lambda\right)\bar{u}_j\left(\frac{1}{2}{\bm P}-{\bm p},\lambda^{\prime}\right) \Gamma_{ij}^{ex} 
  u_j\left(\frac{1}{2}{\bm P}-{\bm p},\lambda^{\prime}\right)u_i\left(\frac{1}{2}{\bm P}+{\bm p},\lambda\right)\right] \ . \label{eq:Gamma2}
\end{eqnarray}
The calculation of the potential part, Eq. (\ref{eq:Vij}), is explained in Appendix \ref{sec:Potential_energy}.

%%%%%%%%%%%%%%%%%%%%%%%%%%%%%%%%%%%%%%%%%%%%%%%%%%
\section{Numerical results \label{sec:numerical_result}}

We present our results of the DBHF calculation, in which we use the Bonn potentials for the NN interaction \cite{Brockmann} and the subtracted T-matrix representation 
explained in Appendix \ref{sec:Self_energy_S0V} (see also Ref. \cite{Gross}).  There are three kinds of 
Bonn potentials, namely Bonn A, B and C, where the NN potential is generated by the exchanges of $\sigma$, $\delta$, $\eta$, $\pi$, $\omega$ and $\rho$ mesons, and 
is parametrized so as to reproduce the deuteron properties and the phase shifts of NN scattering \cite{Brockmann,machleidt}.
The difference among the three potentials is primarily caused by the strength of the tensor force.  

In this paper, our calculation mainly follows the method given in Ref. \cite{Horowitz}.  However,  
we use the subtracted T-matrix representation \cite{Gross} in the self-energy calculation.  
As we will see below, the present results are, however, slightly different from those in Ref. \cite{Gross}, even though we employ the same subtracted 
T-matrix representation as in Ref. \cite{Gross}.  
The reason for it is that, in the calculation of the energy density at the matter-rest frame, we directly use the matrix elements of $\Gamma$ given in Eq. (\ref{eq:Gamma2}),  
which remain unchanged under Lorentz transformation, 
while, in Ref. \cite{Gross}, the energy density is calculated by Eq. (29) in Ref. \cite{Sehn}, in which, instead of $\Gamma$, the self-energies at the rest frame 
are used.  
However, in spite of this difference, the result in Ref. \cite{Gross} is very close to the present one up to high densities.  

Because various results for nuclear matter given by the DBHF calculation have already been reported elsewhere \cite{Horowitz,Haar,Brockmann,Sehn,Gross,Dalen,Klahan,Krastev}, 
we here focus on two topics: (1) the validity of the angle-averaged approximation and the averaging of the total momentum squared of the interacting two nucleons in matter,  
(2) parametrizations for the nucleon self-energies and the EoSs for symmetric nuclear matter and pure neutron matter.

%%%%%%%%%%%%%%%%%%%%%%%%%%%%%%%%%%%%%%%%%%%%%%%%%%
\subsection{Properties of nuclear matter in the DBHF approach\label{sec:nuclear_matter_properties}}

\begin{figure}
\includegraphics[width=250pt,keepaspectratio,clip,angle=270]{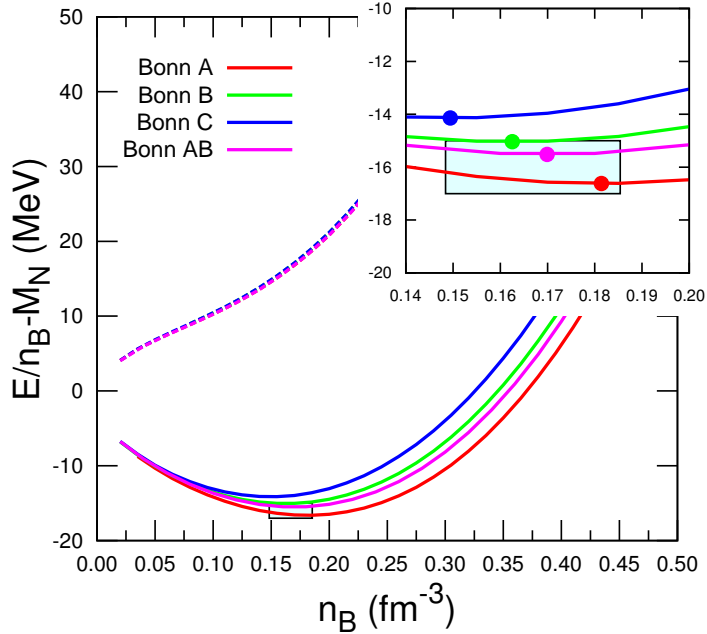}%
\caption{\label{fig:saturation}
 (Color online) Binding energy per particle in symmetric nuclear matter and pure neutron matter as a function of the total baryon density, $n_B$. 
The results for Bonn potentials A, B, C and AB are shown (for details, see the text).  The upper-right, small window shows the magnified part around $n_B^0$. }
\end{figure}

Before we present the main result, we summarize our results for the properties of nuclear matter.  
In Fig. \ref{fig:saturation}, we show the binding energies per particle for symmetric nuclear matter and pure neutron matter in the vicinity of saturation density, $n^0_B$.
In the figure, the rectangular area shows the region of saturation point for symmetric nuclear matter suggested by various semi-empirical data \cite{Bethe_Spring}, that is 
${\cal E}/n_B^0-M_N=-16\pm1$ MeV, and the Fermi momentum, $k_F^0 = 1.35 \pm 0.05$ fm$^{-1}$.

We can see that the saturation points calculated by Bonn A and B are in the rectangular region (see also 
Table \ref{tab:matter_property}, in which we list the several properties of symmetric nuclear matter at the saturation points).  
On the other hand, the result of Bonn C is located outside the region.  It may implies that 
the strength of the tensor force in the interaction strongly relates to the position of the saturation on the Coester band.  
Therefore, it might be interesting to take the averaged values of parameters for the Bonn A and B potentials (which we call Bonn AB) and see how such parameters give 
the saturation properties.  The result of Bonn AB is also shown in Fig. \ref{fig:saturation} and Table \ref{tab:matter_property}.  
From the figure, as expected, we can see that the resulting saturation point lies in between those of Bonn A and B. 

For the case of pure neutron matter, the binding energies for Bonn A, B, C and AB are very close one another, because the tensor force 
mainly influences the ${}^3S_1-{}^3D_1$ coupled states in the two-nucleon system with total isospin $T=1$, and thus it gives less impact on pure neutron matter \cite{Li}.  

\begin{table}
\caption{\label{tab:matter_property}
Calculated properties of symmetric nuclear matter at the saturation points.  We use the Bonn potentials A, B, C and AB (for details, see the text).
The values of the binding energy per particle, ${\cal E}/n_B^0-M_N$, the incompressibility, $K$, the symmetry energy, $S$, the slope parameter, $L$, and 
the effective mass, $M_N^{\ast}$ are in MeV, and the Fermi momentum, $k_F^0$ is in fm$^{-1}$.}
\begin{ruledtabular}
\begin{tabular}{ccccccc}
Potential&$k^0_F$&${\cal E}/n_B^0-M_N$&$K$&$S$&$L$&$M_N^{\ast}$\\
\hline
A&1.39&-16.62&233&34.8&71.2&636\\
B&1.34&-15.04&190&31.2&55.9&666\\
C&1.30&-14.14&170&28.9&46.7&687\\
AB&1.36&-15.52&204&32.3&61.1&653\\
\end{tabular}
\end{ruledtabular}
\end{table}
We comment on the properties of symmetric nuclear matter.  In MFT, it is indispensable to include nonlinear $\sigma$-meson terms phenomenologically to obtain the correct value of 
the nuclear incompressibility, $K = 210 \pm 30$ MeV \cite{Blaizot}.  In contrast, it is not necessary to consider such nonlinear terms in the DBHF approach.  It is remarkable that, 
in the Bonn A, the incompressibility, $K$, is estimated to be $233$ MeV.   However, the symmetry energy, $S$, and the slope parameter, $L$, in Bonn A 
are slightly larger than the values 
suggested in Ref. \cite{Lattimer}.  This may be caused by the fact that the saturation point in Bonn A is located at the relatively deeper binding energy and 
higher density in the allowed, rectangular region.  
We note that the present DBHF calculation can provide the similar results of the density dependences of symmetry energy and the pressure for 
symmetric matter to those calculated in Ref. \cite{Dalen,Klahan}.  

Hereafter, we shall use the Bonn A potential as the optimum interaction. 

%%%%%%%%%%%%%%%%%%%%%%%%%%%%%%%%%%%%%%%%%%%%%%%%%%
\subsection{Dimension-reduction approximation for multi-dimensional integrations\label{sec:dimension_reduction}}

To perform a realistic calculation of the EoS for neutron stars, it is necessary to take into account the N-hyperon (Y) and YY interactions as well as 
the NN interaction, in which we have to consider a lot of two-baryon channels.  Because such calculations should be carried out in a wide range of nuclear 
densities, for example, $n_B = 0 \sim1.2~\mathrm{fm}^{-3}$, we need an enormous amount of calculation time.  
As an attempt to reduce calculation time, 
we study a dimension-reduction approximation for multi-dimensional integrations in the calculations of self-energies and energy density.  
 
In the self-energy calculation, the components, $\Sigma^S, \Sigma^0$ and $\Sigma^V$, are calculated by 
two-dimensional integral equations (see Eqs. (\ref{eq:SigmaS_ps}) - (\ref{eq:SigmaV_cpv}) in Appendix \ref{sec:Self_energy_S0V}). 
We here try to reduce those to one-dimensional ones by exploiting the %trajectory
angle-averaged approximation (see Appendix \ref{sec:Self_energy_S0V} and Ref. \cite{Horowitz}).
We respectively call the self-energies calculated by the two-dimensional integral equations and those by the one-dimensional ones $\Sigma2d$ and $\Sigma1d$.  

\begin{figure}
\includegraphics[width=250pt,keepaspectratio,clip,angle=270]{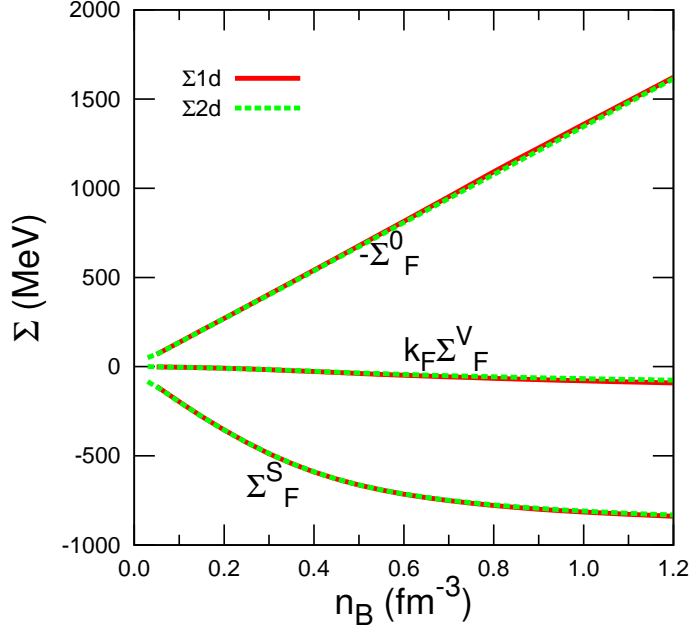}%
\caption{\label{fig:self_energy_1dvs2d}  (Color online) 
Three components of the self-energies at $k = k_{F}$ in symmetric nuclear matter.  The Bonn A potential is used.  
The dashed line is for the two-dimensional calculation ($\Sigma2d$), while the solid line is for the one-dimensional one ($\Sigma1d$). 
}
\end{figure}

In Fig. \ref{fig:self_energy_1dvs2d}, we show the results of $\Sigma1d$ and $\Sigma2d$ in symmetric nuclear matter   
as a function of the total baryon density, $n_B$.
As seen in Fig. \ref{fig:self_energy_1dvs2d}, the curves of $\Sigma1d$ agree very well with those of $\Sigma2d$, i.e. the full calculations, 
over a wide range of density, which implies that the angle-averaged approximation in the self-energy works very well.  
As a consequence, we can replace $\Sigma2d$ by $\Sigma1d$ not only at low densities but also at high densities. 

The potential energy, $\left<\hat{\cal V}_{ij}\right>$, in the energy density is calculated by the three-dimensional integral  
(Eq. (\ref{eq:Vij3}) in Appendix \ref{sec:Potential_energy}), and thus, it is necessary to somehow 
approximate it.   By assuming the angle-averaged approximation, we can again reduce the three-dimensional integral  
to two-dimensional one.  We call the result with this assumption $E2d$, while we call the result of the full calculation $E3d$.  In addition, if we adopt 
the averaging of the total momentum squared of the interacting two nucleons in matter, the two-dimensional integral is further reduced to one-dimensional one  
(see Appendix \ref{sec:Potential_energy}).  We call this result $E1d$. 

\begin{figure}
\includegraphics[width=250pt,keepaspectratio,clip,angle=270]{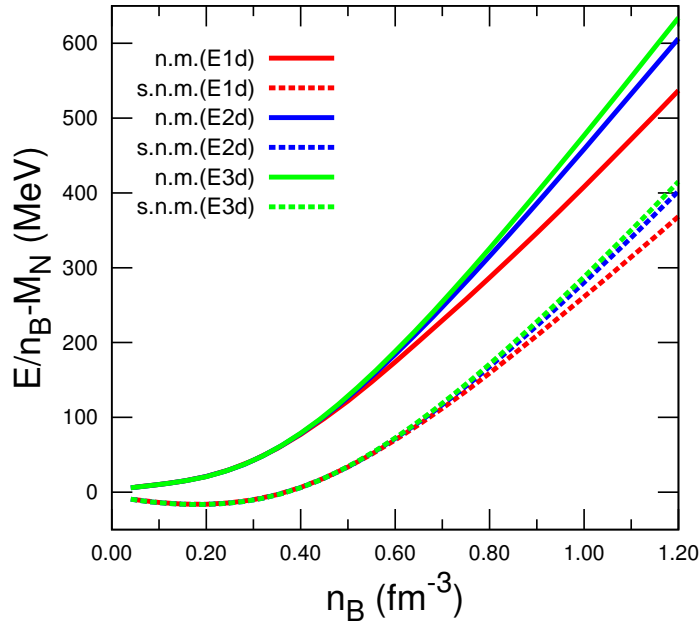}%
\caption{\label{fig:B_123_dim}  (Color online) 
Binding energy per particle for nuclear matter as a function of the total baryon density, $n_B$.  The Bonn A potential and the self-energies, $\Sigma1d$, are used in the 
calculation.  The solid  and dashed curves are, respectively, for the results for pure neutron matter (n.m.) and symmetric nuclear matter (s.n.m.).  
We present the full result, $E3d$, as well as the approximated ones, $E1d$ and $E2d$.}
\end{figure}
In Fig. \ref{fig:B_123_dim}, we compare the full result, $E3d$, with the approximated ones, $E1d$ and $E2d$.  
The three solid (or dashed) curves are in good agreement below $n_B \simeq 0.5~\mathrm{fm}^{-3}$.  
However, increasing the total baryon density, increases the discrepancy between $E3d$ and $E1d$, which implies that the full calculation cannot be approximated by $E1d$ 
at high densities.  In contrast, the difference between $E2d$ and $E3d$ is still rather small even at high densities. 

\begin{figure}
\includegraphics[width=250pt,keepaspectratio,clip,angle=270]{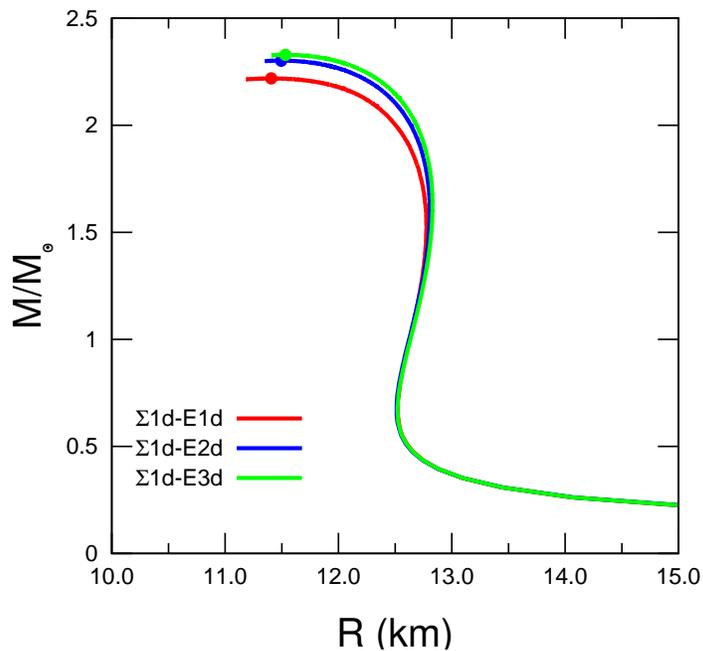}%
\caption{\label{fig:Stellar_BPS_BBP}  (Color online)  
Neutron-star mass versus radius in three different approximations. The Bonn A potential is used. 
}
\end{figure}
Next, using $\Sigma1d$, we calculate the neutron-star mass as a function of the radius, which is given in Fig. \ref{fig:Stellar_BPS_BBP}.  We also 
list the properties of neutron stars in Table \ref{tab:MR}. 
The present calculation is restricted to the neutron-star matter composed of protons, neutrons, electrons and muons.  Under the conditions of charge neutrality 
and $\beta$-equilibrium in weak interaction, we solve the TOV equation \cite{TOV} to obtain the mass-radius relation.  
So far, in the DBHF calculation, the EoS for neutron-star matter is calculated by using the parabolic interpolation between the EoSs for symmetric nuclear matter 
and pure neutron matter (see, for example, Ref. \cite{Klahan,Krastev}).  However, in the present work, we calculate the EoS for asymmetric nuclear matter without 
such interpolations.
\begin{table}
\caption{\label{tab:MR}
Neutron-star radius, $R_{max}$ (in km), the central density, $n_c$ (in fm$^{-3}$), and the ratio of the maximum neutron-star mass to the solar mass, $M_{max}/M_{\odot}$. 
The Bonn A potential is used. }
\begin{ruledtabular}
\begin{tabular}{lccc}
Case&$R_{max}$&$n_c$&$M_{max}/M_{\odot}$\\
\hline
$\Sigma1d-E1d$&11.4&0.95&2.22\\
$\Sigma1d-E2d$&11.5&0.92&2.30\\
$\Sigma1d-E3d$&11.5&0.91&2.33\\
\end{tabular}
\end{ruledtabular}
\end{table}

As expected from the result shown in Fig. \ref{fig:B_123_dim}, the EoS by $\Sigma1d - E3d$ gives the heaviest mass, $2.33~M_{\odot}$, and the large difference 
between the maximum masses by $\Sigma1d - E3d$ and $\Sigma1d - E1d$ is seen in Fig. \ref{fig:Stellar_BPS_BBP}.  
However, the mass by $\Sigma1d - E2d$ is very close to that by $\Sigma1d - E3d$.  

Thus, we can conclude that the combination of approximations, $\Sigma1d$ and $E2d$, to the calculations of self-energies and energy density 
is the most suitable one to save calculation time.   Furthermore, we may be able to expected that the difference between $E3d$ and $E1d$ becomes smaller in 
the neutron star containing hyperons as well as nucleons, because the creation of hyperons reduces the fraction of nucleon density in matter, and thus 
the high density part in Fig. \ref{fig:B_123_dim} may not contribute much to the EoS in such cases. 

%%%%%%%%%%%%%%%%%%%%%%%%%%%%%%%%%%%%%%%%%%%%%%%%%%
\subsection{Parametrizations\label{sec:parametrization}} 

For convenient use of the DBHF result, we provide useful parametrizations for the self-energy components 
and the EoS for nuclear matter.  

Each component of the nucleon self-energy, which depends on the momentum, $k$, and the Fermi momentum, $k_{F i}$ ($i = p$ or $n$), can be parametrized 
by the following, power-series expansion: 
\begin{eqnarray}
	&&\Sigma_n^S(k_{Fn},k_{Fp},k)\nonumber\\
		&=&-1033 + 2284k_{Fn} - 1767k_{Fn}^2 + 425.2k_{Fn}^3 - 29.29k_{Fn}^4 + 384.9k_{Fp} - 386.3k_{Fn}k_{Fp}\nonumber\\
  	&&+ 148.3k_{Fn}^2k_{Fp} - 25.01k_{Fn}^3k_{Fp} -  201.2k_{Fp}^2 + 94.37k_{Fn}k_{Fp}^2 - 5.789k_{Fn}^2k_{Fp}^2 - 32.46k_{Fp}^3\nonumber\\
  	&&-  2.635k_{Fn}k_{Fp}^3 + 12.54k_{Fp}^4 + 172.1k - 267.1k_{Fn}k +  102.0k_{Fn}^2k - 10.51k_{Fn}^3k - 71.29k_{Fp}k\nonumber\\
  	&&+ 10.60k_{Fn}k_{Fp}k +  16.66k_{Fn}^2k_{Fp}k + 36.53k_{Fp}^2k - 16.82k_{Fn}k_{Fp}^2k - 3.828k_{Fp}^3k -  52.33k^2\nonumber\\ 
  	&&+ 136.6k_{Fn}k^2 - 31.39k_{Fn}^2k^2 + 25.97k_{Fp}k^2-  16.83k_{Fn}k_{Fp}k^2 + 7.546k_{Fp}^2k^2 - 45.74k^3\nonumber\\
  	&&+ 2.287k_{Fn}k^3 -  0.8569k_{Fp}k^3 + 6.430k^4 \ ,
\end{eqnarray}
\begin{eqnarray}
	&&\Sigma_p^S(k_{Fn},k_{Fp},k)\nonumber\\
		&=&-893.2 + 1869k_{Fn} - 1423k_{Fn}^2 + 380.5k_{Fn}^3 - 32.92k_{Fn}^4 + 545.7k_{Fp} - 687.3k_{Fn}k_{Fp}\nonumber\\
    &&+ 239.8k_{Fn}^2k_{Fp} - 28.93k_{Fn}^3k_{Fp} - 107.3k_{Fp}^2 + 224.8k_{Fn}k_{Fp}^2 - 27.68k_{Fn}^2k_{Fp}^2 - 231.1k_{Fp}^3\nonumber\\
    &&-  21.26k_{Fn}k_{Fp}^3 + 60.10k_{Fp}^4 + 54.78k + 13.65k_{Fn}k +  19.75k_{Fn}^2k - 0.4057k_{Fn}^3k - 105.2k_{Fp}k\nonumber\\
    &&- 56.77k_{Fn}k_{Fp}k - 11.83k_{Fn}^2k_{Fp}k + 68.64k_{Fp}^2k + 30.57k_{Fn}k_{Fp}^2k - 19.26k_{Fp}^3k -  87.59k^2\nonumber\\
    &&+ 24.20k_{Fn}k^2 - 7.972k_{Fn}^2k^2 + 86.62k_{Fp}k^2 +  7.563k_{Fn}k_{Fp}k^2 - 8.808k_{Fp}^2k^2 + 7.976k^3\nonumber\\
    &&+ 0.7710k_{Fn}k^3 -  25.90k_{Fp}k^3 + 7.593k^4 \ ,
\end{eqnarray}
\begin{eqnarray}
	&&\Sigma_n^0(k_{Fn},k_{Fp},k)\nonumber\\
		&=&191.6 - 573.1k_{Fn} + 515.9k_{Fn}^2 - 245.7k_{Fn}^3 + 17.55k_{Fn}^4 - 88.93k_{Fp} + 111.9k_{Fn}k_{Fp}\nonumber\\
    &&- 45.05k_{Fn}^2k_{Fp} - 1.718k_{Fn}^3k_{Fp} + 113.6k_{Fp}^2 - 58.68k_{Fn}k_{Fp}^2 + 21.11k_{Fn}^2k_{Fp}^2 - 56.05k_{Fp}^3\nonumber\\
    &&- 4.129k_{Fn}k_{Fp}^3 + 1.502k_{Fp}^4 + 202.5k - 203.2k_{Fn}k - 0.6211k_{Fn}^2k + 17.24k_{Fn}^3k - 45.30k_{Fp}k\nonumber\\
    &&+ 8.512k_{Fn}k_{Fp}k + 19.11k_{Fn}^2k_{Fp}k + 19.44k_{Fp}^2k - 22.48k_{Fn}k_{Fp}^2k + 3.111k_{Fp}^3k - 112.8k^2\nonumber\\
    &&+ 174.3k_{Fn}k^2 - 22.86k_{Fn}^2k^2 + 10.78k_{Fp}k^2 - 15.74k_{Fn}k_{Fp}k^2 + 9.332k_{Fp}^2k^2 - 28.79k^3\nonumber\\
    &&- 19.35k_{Fn}k^3 +  1.244k_{Fp}k^3 + 12.31k^4 \ ,
\end{eqnarray}
\begin{eqnarray}
	&&\Sigma_p^0(k_{Fn},k_{Fp},k)\nonumber\\ 
		&=&-37.78 + 84.47k_{Fn} + 30.83k_{Fn}^2 - 75.99k_{Fn}^3 + 6.714k_{Fn}^4 -  31.29k_{Fp} - 60.14k_{Fn}k_{Fp}\nonumber\\
    &&+ 38.01k_{Fn}^2k_{Fp} - 10.13k_{Fn}^3k_{Fp} +  19.28k_{Fp}^2 + 31.65k_{Fn}k_{Fp}^2 + 10.98k_{Fn}^2k_{Fp}^2 - 79.14k_{Fp}^3\nonumber\\
    &&- 27.68k_{Fn}k_{Fp}^3 + 6.778k_{Fp}^4 - 8.749k + 44.05k_{Fn}k - 0.004370k_{Fn}^2k + 0.01541k_{Fn}^3k\nonumber\\
    &&+ 25.79k_{Fp}k - 47.72k_{Fn}k_{Fp}k + 1.585k_{Fn}^2k_{Fp}k - 36.53k_{Fp}^2k + 8.365k_{Fn}k_{Fp}^2k + 13.95k_{Fp}^3k\nonumber\\
    &&- 56.79k^2 - 15.44k_{Fn}k^2 + 1.677k_{Fn}^2k^2 + 49.77k_{Fp}k^2 + 9.450k_{Fn}k_{Fp}k^2 + 10.32k_{Fp}^2k^2\nonumber\\
    &&+ 28.56k^3 - 0.02399k_{Fn}k^3 - 37.12k_{Fp}k^3 + 9.125k^4 \ ,
\end{eqnarray}
\begin{eqnarray}
	&&\Sigma_n^V(k_{Fn},k_{Fp},k)\nonumber\\ 
		&=&0.1521 - 0.5998k_{Fn} + 0.8179k_{Fn}^2 - 0.4307k_{Fn}^3 + 0.06665k_{Fn}^4 - 0.2702k_{Fp} + 0.2533k_{Fn}k_{Fp}\nonumber\\
    &&- 0.1394k_{Fn}^2k_{Fp} + 0.03638k_{Fn}^3k_{Fp} + 0.3053k_{Fp}^2 + 0.0007945k_{Fn}k_{Fp}^2 - 0.001880k_{Fn}^2k_{Fp}^2\nonumber\\
    &&- 0.2312k_{Fp}^3 - 0.01501k_{Fn}k_{Fp}^3 + 0.05148k_{Fp}^4 + 0.3278k - 0.6618k_{Fn}k + 0.4069k_{Fn}^2k\nonumber\\
    &&- 0.07270k_{Fn}^3k - 0.05230k_{Fp}k + 0.03404k_{Fn}k_{Fp}k - 0.05118k_{Fn}^2k_{Fp}k + 0.02249k_{Fp}^2k\nonumber\\
    &&+ 0.05848k_{Fn}k_{Fp}^2k - 0.02161k_{Fp}^3k + 0.08527k^2 - 0.1199k_{Fn}k^2 + 0.03291k_{Fn}^2k^2\nonumber\\
    &&+ 0.0006560k_{Fp}k^2 + 0.02024k_{Fn}k_{Fp}k^2 - 0.01927k_{Fp}^2k^2 + 0.02552k^3 - 0.006904k_{Fn}k^3\nonumber\\
    &&- 0.0008859k_{Fp}k^3 - 0.002613k^4 \ ,
\end{eqnarray}
\begin{eqnarray}
	&&\Sigma_p^V(k_{Fn},k_{Fp},k)\nonumber\\
		&=&0.1887 - 0.1975k_{Fn} + 0.1287k_{Fn}^2 - 0.08467k_{Fn}^3 + 0.01424k_{Fn}^4 - 0.6801k_{Fp} + 0.5139k_{Fn}k_{Fp}\nonumber\\
    &&- 0.1847k_{Fn}^2k_{Fp} + 0.02296k_{Fn}^3k_{Fp} + 0.6572k_{Fp}^2 - 0.1481k_{Fn}k_{Fp}^2 + 0.02273k_{Fn}^2k_{Fp}^2 - 0.3207k_{Fp}^3\nonumber\\
    &&+ 0.01080k_{Fn}k_{Fp}^3 + 0.05429k_{Fp}^4 + 0.1452k - 0.08241k_{Fn}k + 0.07264k_{Fn}^2k - 0.01388k_{Fn}^3k\nonumber\\
    &&- 0.1953k_{Fp}k - 0.06356k_{Fn}k_{Fp}k + 0.001918k_{Fn}^2k_{Fp}k + 0.1436k_{Fp}^2k + 0.02130k_{Fn}k_{Fp}^2k\nonumber\\
    &&- 0.03376k_{Fp}^3k - 0.05190k^2 + 0.01251k_{Fn}k^2 + 0.003157k_{Fn}^2k^2 + 0.05247k_{Fp}k^2\nonumber\\
    &&- 0.006964k_{Fn}k_{Fp}k^2 - 0.007900k_{Fp}^2k^2 - 0.004147k^3 - 0.002807k_{Fn}k^3 + 0.002096k_{Fp}k^3\nonumber\\
    &&- 0.0006445k^4 \ . 
\end{eqnarray}
These are fitted so as to reproduce the results of the full ($\Sigma2d - E3d$) calculation with Bonn A, and valid for the region of 
$0.03~\mathrm{fm}^{-3}\le n_B\le1.2~\mathrm{fm}^{-3}$, $0\le Y_p\le0.5$ and $ k \le k_{Fn} \, (k_{Fp})$ for $\Sigma_n \, (\Sigma_p)$, 
where $Y_p$ is the proton fraction, $Y_p\equiv n_p/n_B$, with $n_p$ being the proton density.  

\begin{figure}
\includegraphics[width=250pt,keepaspectratio,clip,angle=270]{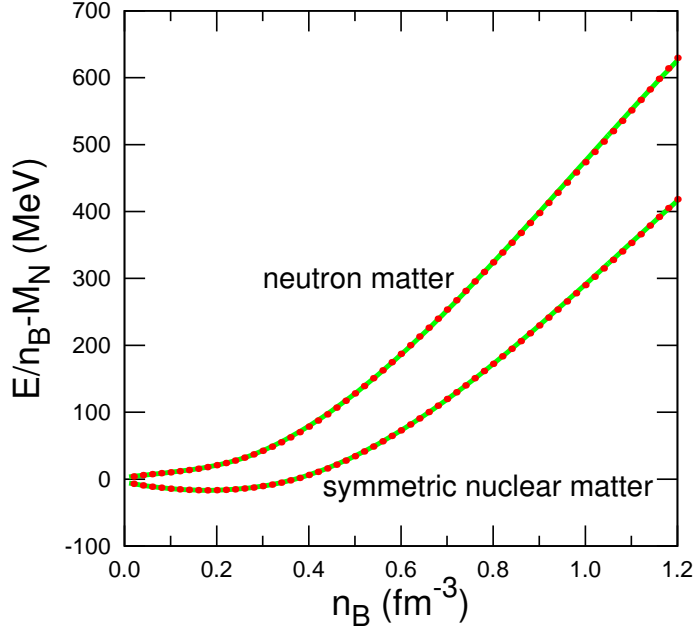}%
\caption{\label{fig:fit_B} (Color online) 
EoS for nuclear matter as a function of the total baryon density, $n_B$.  
The DBHF results are represented by the (red) dotted curves, while the parametrizations are denoted by the (green) solid curves.}
\end{figure}

In addition, we also provide parametrizations for the EoSs for symmetric nuclear matter (s.n.m.) and pure neutron matter (n.m.) in terms of 
polynomials of the nuclear density: 
\begin{eqnarray}
 \left(\frac{\cal E}{n_B}-M_N\right)_{s.n.m.}  &=& 
	\left\{
	\begin{array}{cc}
		-4.570-130.4n_B+326.1n_B^2+144.5n_B^3& \  \mathrm{for} \ 0 < n_B \le 0.3\\
		25.11-366.1n_B+908.3n_B^2-276.3n_B^3& \ \ \mathrm{for} \ 0.3 <n_B \le 1.2 
	\end{array}
	\right. \ , \nonumber\\
	\label{eq:binding_energy_snm_fit}\\
	\left(\frac{\cal E}{n_B}-M_N\right)_{n.m.}&=&
	\left\{
	\begin{array}{cc}
		2.453+94.62n_B-304.6n_B^2+1449n_B^3&  \  \mathrm{for} \ 0 <n_B \le 0.3 \\
		22.57-200.3n_B+991.0n_B^2-337.6n_B^3&  \ \  \mathrm{for} \ 0.3 < n_B \le1.2 
	\end{array}
	\right. \ , \nonumber\\
	\label{eq:binding_energy_nm_fit}
\end{eqnarray}
where $n_B$ is in fm$^{-3}$. 
These are again fitted to the results of the full calculation with Bonn A. 
 In Eqs. (\ref{eq:binding_energy_snm_fit}) and (\ref{eq:binding_energy_nm_fit}), the density region is divided into two parts at $n_B = 0.3$ fm$^{-3}$, 
and, at that density, the value of polynomial function and its derivative are respectively continuous.  
These parametrizations are shown in Fig. \ref{fig:fit_B}, and they work very well up to very high density.   

\begin{figure}
\includegraphics[width=250pt,keepaspectratio,clip,angle=270]{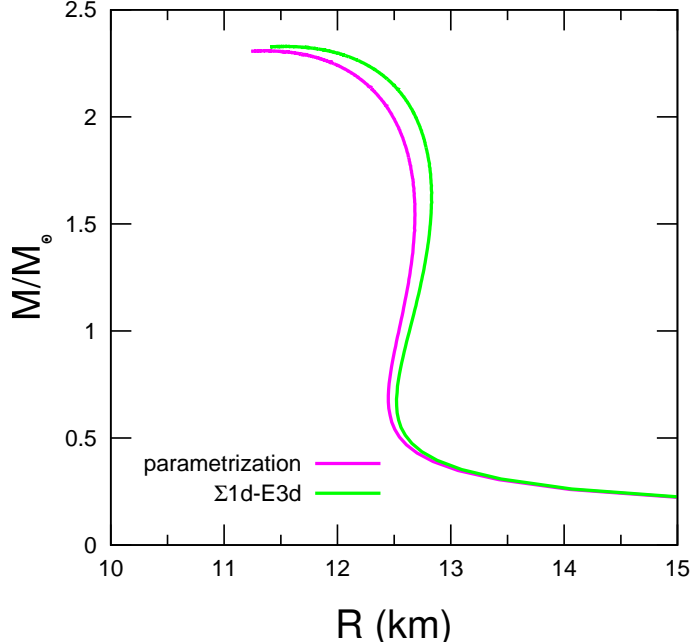}%
\caption{\label{fig:parabolic_interpolation} (Color online) 
Neutron-star mass versus radius with or without the parabolic interpolation. 
The green curve is for the result by $\Sigma1d-E3d$ (without the interpolation), while the pink curve is for that by the interpolation between 
Eqs. (\ref{eq:binding_energy_snm_fit}) and (\ref{eq:binding_energy_nm_fit}).}
\end{figure}

For the calculation of asymmetric nuclear matter, it may practically be convenient to use the parabolic interpolation between 
Eqs. (\ref{eq:binding_energy_snm_fit}) and (\ref{eq:binding_energy_nm_fit}), because the difference between the results by the interpolation and 
$\Sigma1d-E3d$ is not large (see Fig. \ref{fig:parabolic_interpolation}). 

%%%%%%%%%%%%%%Summary and Conclusions%%%%%%%%%%%%%
\section{Summary and conclusions\label{sec:summary_and_conclusion}}

Using the DBHF approximation, we have investigated the properties of isospin-asymmetric nuclear matter not only in the vicinity of saturation density 
but also for high densities, and have applied the DBHF approach to neutron stars.  

We have first confirmed that the present DBHF approach, in which the subtracted T-matrix representation is applied to the Bethe-Salpeter equations with the Bonn potentials, 
can mostly explain the observed, saturation properties of symmetric nuclear matter.  
Next, since the DBHF calculation is very complicated and needs enormous calculation time, to handle it as easily as possible, 
we have examined a dimension-reduction approximation, 
namely the angle-averaged approximation and the averaging of the total momentum squared of the interacting two nucleons in matter, 
in the calculations of nucleon self-energies and energy density of matter. 

By applying the angle-averaged approximation, the self-energy components can be calculated very accurately by the one-dimensional integral equations up to high densities  
($\sim 1.2~\mathrm{fm}^{-3}$).  However, in the calculation of energy density, which requires the three-dimensional integral, 
the validity of the dimension-reduction approximation depends on the total baryon density, 
that is, by applying both the angle-averaged approximation and the averaging of the total momentum squared of two nucleons, the three-dimensional integral  
can be well approximated by the one-dimensional one in the vicinity of saturation density, $n_B^0$, while, at high densities, the averaging of the total momentum 
squared does not work well, and thus we must at least handle the two-dimensional integral. 

In the present paper, we have provided simple and useful parametrizations for the nucleon self-energies and the EoSs for symmetric nuclear matter and 
pure neutron matter, which are fitted so as to reproduce the results of the full DBHF calculation and cover the density range of $0 \sim 1.2~\mathrm{fm}^{-3}$. 

Finally, we comment on future works. 
We have not considered the degrees of freedom of hyperons in the present calculation.  It is generally believed that the inclusion of hyperons in neutron stars inevitably softens 
the EoS and thereby the maximum mass of neutron star is seriously reduced.  However, it might be premature to conclude so, 
because, within relativistic Hartree-Fock approximation, the creation of hyperons in neutron stars is strongly suppressed and thus the EoS is not so soften \cite{katayama}. 
In any case, it is very vital to perform the full DBHF calculation including hyperons as well as nucleons to address this issue.  The dimension-reduction approximation we have 
proposed in the present paper may help us accomplish such heavy DBHF calculations.  

It is also important to consider the $K$- and $K^{\ast}$-meson exchanges between two baryons in matter.  In MFT, they do not contribute to the EoS because of 
the absence of Fock diagrams. 
When the Fock terms are taken into account, the $K$- and $K^{\ast}$-meson exchanges mix some baryons in matter.  
The $\pi$ and $\rho$ mesons also mix the $\Sigma$ and $\Lambda$ through the Fock diagrams.  

At very high density, the quark and gluon degrees of freedom, rather than the hadron degrees of freedom, take place in neutron stars \cite{Glen}.  Thus, it is eventually 
necessary to take into account such degrees of freedom and consider the phase transition between the quark-gluon and hadron phases.  

\vspace{1cm}
\begin{acknowledgements}
This work was supported by JSPS KAKENHI Grant Number 255742.  We thank Tomoyuki Maruyama and N. Yasutake for valuable suggestions. 
\end{acknowledgements}

%%%%%%%%%%%%%%Appendix%%%%%%%%%%%%%%%%%%%%%%%%%%%
\appendix
\section{Self-energy components \label{sec:Self_energy_S0V}}

The effective interactions, $\Gamma$, in the ladder-approximated Bethe-Salpeter equations, Eqs. (\ref{eq:Gammaii})-(\ref{eq:Gammaijex}), are most conveniently 
calculated in the center-of-mass frame of the interacting two nucleons.  
In contrast, using the $\Gamma$, the self-energies, Eq. (\ref{eq:Self_energy}), must be calculated in the matter-rest frame.
Therefore, the effective interactions, $\Gamma$, have to be transformed between the two frames.   

The present calculation of the self-energies follows the works by Faessler's group \cite{Gross,Dalen}, where they have proposed 
the subtracted T-matrix representation which is composed of pseudoscalar (ps) and complete pseudovector (pv) representations. 
The complete pv representation is then applied to the effective interaction due to the {\it single} $\pi$- and $\eta$-meson exchanges, while 
the ps representation is used for the {\it multiple} $\pi$- and $\eta$-meson exchanges and the other heavy-meson exchanges.  

The effective interaction is divided into five Lorentz-covariants, that is,  
the scalar ($S$), vector ($V$), tensor ($T$), axial-vector ($A$) and pseudoscalar ($P$) components in the ps representation, while, instead of $P$, 
the pseudovector ($PV$) component takes place in the pv representation.  
Moreover, by setting a different type of pv representation, one can get rid of undesired ps contributions due to the $\pi$-meson exchange.  
This new one is designated as the complete pv representation \cite{Gross}.  

In the ps representation, the components of on-shell self-energy are given by
\begin{eqnarray}
	&&\Sigma_i^S(\underline{k},E^{\ast}_i(k)) \nonumber \\
	&=&\sum_{j=p,n}\int\frac{d^3q}{(2\pi)^3}\frac{M^{\ast}_j}{E_j^{\ast}(q)}\theta(k_{Fj}-|{\bm q}|)\left(2{}^S\Gamma_{ij}^{dir}-	\frac{1}{2}\sum_m\xi_{ms}{}^{m}\Gamma_{ij}^{ex}\right) \ , 
 \label{eq:SigmaS_ps} \\ 
	&&\Sigma_i^0(\underline{k},E^{\ast}_i(k)) \nonumber \\
	&=&-\sum_{j=p,n}\int\frac{d^3q}{(2\pi)^3}\theta(k_{Fj}-|{\bm q}|)\left(2{}^V\Gamma_{ij}^{dir}-\frac{1}{2}\sum_m\xi_{mv}{}^{m}\Gamma_{ij}^{ex}\right) \ , \label{eq:Sigma0_ps} \\
	&&\Sigma_i^V(\underline{k},E^{\ast}_i(k)) \nonumber \\
	&=&-\sum_{j=p,n}\frac{1}{\underline{k}^2}\int\frac{d^3q}{(2\pi)^3}\frac{{\bm q}\cdot{\bm k}}{E_j^{\ast}(q)}\theta(k_{Fj}-|{\bm q}|)\left(2{}^V\Gamma_{ij}^{dir}
 -\frac{1}{2}\sum_m\xi_{mv}{}^{m}\Gamma_{ij}^{ex}\right) \ , \label{eq:SigmaV_ps}
\end{eqnarray}
with $\xi_{ms}$ and $\xi_{mv}$ being the exchange coefficients (see Table \ref{tab:Exchange_coefficients}).  
In the complete pv representation, they are written as
\begin{eqnarray}
	&&\Sigma_i^S(\underline{k},E^{\ast}_i(k)) \nonumber \\
	&=&\sum_{j=p,n}\int\frac{d^3q}{(2\pi)^3}\theta(k_{Fj}-|{\bm q}|)\frac{M_j^{\ast}}{2E_j^{\ast}(q)}\left[4\left({}^Sg_{ij}^{dir}+{}^{\tilde{S}}g_{ij}^{ex}\right)
 -\left({}^Sg_{ij}^{ex}+{}^{\tilde{S}}g_{ij}^{dir}\right)+4\left({}^Ag_{ij}^{ex}+{}^{A}g_{ij}^{dir}\right)\right. \nonumber \\
	&&\left.-\frac{1}{(M_i^{\ast}+M_j^{\ast})^2}\left\{M_i^{\ast2}+M_j^{\ast2}-2\left(E_i^{\ast}(k)E_j^{\ast}(q)-{\bm k}\cdot{\bm q}\right)\right\} 
 \left({}^{PV}g_{ij}^{ex}+{}^{\tilde{PV}}g_{ij}^{dir}\right)\right] \ , \label{eq:SigmaS_cpv}\\
	&&\Sigma_i^0(\underline{k},E^{\ast}_i(k)) \nonumber \\
	&=&\sum_{j=p,n}\int\frac{d^3q}{(2\pi)^3}\frac{\theta(k_{Fj}-|{\bm q}|)}{2}\left\{\left({}^{S}g_{ij}^{ex}+{}^{\tilde{S}}g_{ij}^{dir}\right)-2\left({}^{A}g_{ij}^{ex}+{}^{A}g_{ij}^{dir}\right)\right. \nonumber \\
	&&+\frac{1}{(M_i^{\ast}+M_j^{\ast})^2E_j^{\ast}(q)}\left[2E_i^{\ast}(k)\left(M_j^{\ast2}-E_j^{\ast}(q)E_i^{\ast}(k)+{\bm q}\cdot{\bm k}\right)\right. \nonumber \\
	&&\left.\left.-E_j^{\ast}(q)\left(M_j^{\ast2}-M_i^{\ast2}\right)\right]\left({}^{PV}g_{ij}^{ex}+{}^{\tilde{PV}}g_{ij}^{dir}\right)\right\} \ ,  \label{eq:Sigma0_cpv} \\
	&&\Sigma_i^V(\underline{k},E^{\ast}_i(k)) \nonumber \\
	&=&\sum_{j=p,n}\frac{1}{\underline{k}^2}\int\frac{d^3q}{(2\pi)^3}\frac{\theta(k_{Fj}-|{\bm q}|)}{2E_j^{\ast}(q)}\left\{{\bm q}\cdot{\bm k} 
 \left[\left({}^{S}g_{ij}^{ex}+{}^{\tilde{S}}g_{ij}^{dir}\right)-2\left({}^{A}g_{ij}^{ex}+{}^{A}g_{ij}^{dir}\right)\right]\right. \nonumber \\
	&&-\frac{1}{(M_i^{\ast}+M_j^{\ast})^2}\left[-2\underline{k}^2\left(M_j^{\ast2}-E_j^{\ast}(q)E_i^{\ast}(k)+{\bm q}\cdot{\bm k}\right)\right. \nonumber \\
	&&\left.\left.-{\bm q}\cdot{\bm k}\left(M_i^{\ast2}-M_j^{\ast2}\right)\right]\left({}^{PV}g_{ij}^{ex}+{}^{\tilde{PV}}g_{ij}^{dir}\right)\right\} \ , \label{eq:SigmaV_cpv}
\end{eqnarray}
with a linear transformation 
\begin{equation}
	\left[
	\begin{array}{c}
		{}^Sg_{ij}^{dir,ex}\\
		{}^{\tilde{S}}g_{ij}^{dir,ex}\\
		{}^{A}g_{ij}^{dir,ex}\\
		{}^{PV}g_{ij}^{dir,ex}\\
		{}^{\tilde{PV}}g_{ij}^{dir,ex}\\
	\end{array}
	\right]=\frac{1}{4}\left[
	\begin{array}{ccccc}
	4&-2&-8&0&-2\\
	0&-6&-16&0&2\\
	0&2&0&0&2\\
	0&2&-8&4&2\\
	0&6&-16&0&-2\\
	\end{array}
	\right]\left[
	\begin{array}{c}
	{}^S\Gamma_{ij}^{dir,ex}\\
	{}^V\Gamma_{ij}^{dir,ex}\\
	{}^T\Gamma_{ij}^{dir,ex}\\
	{}^{PV}\Gamma_{ij}^{dir,ex}\\
	{}^A\Gamma_{ij}^{dir,ex}\\
	\end{array}
	\right] \ . 
\end{equation}
Here, ${\bm k}$ (${\bm q}$) is the three momentum of nucleon $i$ ($j$), 
and $|{\bm k}^{\ast}|=|{\bm k}|\equiv\underline{k}$ ($|{\bm q}^{\ast}|=|{\bm q}|\equiv\underline{q}$) in the matter-rest frame.
The unphysical amplitudes in $\Gamma$, which are not fully anti-symmetrized under the exchange between the interacting two nucleons, 
must be eliminated before adding up $\Gamma$ in Eqs. (\ref{eq:SigmaS_ps})-(\ref{eq:SigmaV_cpv}). 
\begin{table}
\caption{\label{tab:Exchange_coefficients}
Exchange coefficients for the exchange amplitudes.}
\begin{ruledtabular}
\begin{tabular}{ccc}
$m$&$\xi_{ms}$&$\xi_{mv}$\\
\hline
$S$&1&1\\
$V$&4&-2\\
$T$&12&0\\
$P$&1&-1\\
$A$&-4&-2\\
\end{tabular}
\end{ruledtabular}
\end{table}

Because, in the self-energies, the azimuthal integral in spherical coordinates ($\underline{q}, \theta, \phi$) can be performed analytically, the three-dimensional integrals 
are reduced to the two-dimensional ones.  However, because the two-dimensional integrals still require enormous calculation time, the calculation of the self-energies 
is not an easy task.  Hence, it is very useful to find an appropriate approximation to reduce calculation time further. 

In Ref. \cite{Horowitz}, exploiting a trajectory, which is defined by the relative momentum 
\begin{equation}
	{\bm t} \equiv \frac{1}{2}({\bm k}-{\bm q})  \ ,  \label{eq:t_av}
\end{equation}
and, using the angle averaging over the angle $\hat{\bm k} \cdot \hat{\bm t}$ ($\hat{\bm k}$ and $\hat{\bm t}$ being the unit vectors along ${\bm k}$ and ${\bm t}$, 
respectively) in Eqs. (\ref{eq:SigmaS_ps})-(\ref{eq:SigmaV_cpv}), the two-dimensional integrals are reduced to the one-dimensional ones.  
Thus, using this method, the angle-averaged value of the total momentum squared, $P_{av}^2$, and the effective energy, $E_{j \, av}^{\ast}$, can be written as 
\begin{eqnarray}
	P_{av}^2&\equiv&\frac{\int d\Omega_t\, t^2 \theta(k_{Fj}-|2{\bm t}-{\bm k}|)4({\bm k}-{\bm t})^2}{\int d\Omega_t\, t^2 \theta(k_{Fj}-|2{\bm t}-{\bm k}|)}\nonumber\\
	&=&\left\{
	\begin{array}{lc}
	4(\underline{k}^2+t^2),&0\le t\le\frac{1}{2}(k_{Fj}-\underline{k}),~k_{Fj}\geq \underline{k}\\\\
	3\underline{k}^2+k_{Fj}^2-4\underline{k}t,&\frac{1}{2}(k_{Fj}-\underline{k})< t\le\frac{1}{2}(k_{Fj}+\underline{k}),~k_{Fj}\geq \underline{k}\\\\
	3\underline{k}^2+k_{Fj}^2-4\underline{k}t,&\frac{1}{2}(\underline{k}-k_{Fj})< t\le\frac{1}{2}(k_{Fj}+\underline{k}),~k_{Fj}<\underline{k}
	\end{array}
	\right. \ , \label{eq:P_av} \\
	E_{j\,av}^{\ast}(t)&\equiv&\frac{\int d\Omega_t\, t^2 \theta(k_{Fj}-|2{\bm t}-{\bm k}|)E_j^{\ast}(q)}{\int d\Omega_t\, t^2 \theta(k_{Fj}-|2{\bm t}-{\bm k}|)}\nonumber\\
	&=&\left\{
	\begin{array}{lc}
	\frac{[(\underline{k}+2t)^2+M_j^{\ast2}]^{3/2}-[(\underline{k}-2t)^2+M_j^{\ast2}]^{3/2}}{12\underline{k}t},&0\le t\le\frac{1}{2}(k_{Fj}-k),~k_{Fj}\geq \underline{k}\\\\
	\frac{(k_{Fj}^2+M_j^{\ast2})^{3/2}-[(\underline{k}-2t)^2+M_j^{\ast2}]^{3/2}}{\frac{3}{2}[k_{Fj}^2-(\underline{k}-2t)^2]},&\frac{1}{2}(k_{Fj}-\underline{k})< 
 t\le\frac{1}{2}(k_{Fj}+k),~k_{Fj}\geq \underline{k}\\\\
	\frac{(k_{Fj}^2+M_j^{\ast2})^{3/2}-[(\underline{k}-2t)^2+M_j^{\ast2}]^{3/2}}{\frac{3}{2}[k_{Fj}^2-(\underline{k}-2t)^2]},&\frac{1}{2}(\underline{k}-k_{Fj})<  
 t\le\frac{1}{2}(k_{Fj}+k),~k_{Fj}<\underline{k}
	\end{array}
	\right. \ , \nonumber \\ \label{eq:E_av}
\end{eqnarray}
with 
$t = |{\bm t}|$ and $d\Omega_t = \sin\theta \, d\phi d\theta$. 

Substituting Eqs. (\ref{eq:t_av})-(\ref{eq:E_av}) into Eqs. (\ref{eq:SigmaS_ps})-(\ref{eq:SigmaV_ps}) and Eps. (\ref{eq:SigmaS_cpv})-(\ref{eq:SigmaV_cpv}), 
the self-energies in the ps representation read 
\begin{eqnarray}
	&&\Sigma_i^S(\underline{k},E^{\ast}_i(k))\nonumber\\
	&=&\sum_{j=p,n}\theta(k_{Fj})\left[\frac{4}{\pi^2}\int_{0}^{\frac{1}{2}|k_{Fj}-\underline{k}|}\theta(k_{Fj}-\underline{k})dt\,t^2\frac{M^{\ast}_j}{E_{j\,av}^{\ast}(t)}\left(2{}^S\Gamma_{ij}^{dir}
 -\frac{1}{2}\sum_m\xi_{ms}{}^{m}\Gamma_{ij}^{ex}\right)\right.\nonumber\\
	&&+\frac{1}{2\pi^2}\int_{\frac{1}{2}|k_{Fj}-\underline{k}|}^{\frac{1}{2}(k_{Fj}+k)}dt\frac{M^{\ast}_j}{E_{j\,av}^{\ast}(t)}\frac{t}{\underline{k}}(k_{Fj}^2-4t^2-\underline{k}^2
  +4t \underline{k})  \nonumber\\
	&&\times\left.\left(2{}^S\Gamma_{ij}^{dir}-\frac{1}{2}\sum_m\xi_{ms}{}^{m}\Gamma_{ij}^{ex}\right)\right],\\
  &&\Sigma_i^0(\underline{k},E^{\ast}_i(k))\nonumber\\
  &=&\sum_{j=p,n}\theta(k_{Fj})\left[-\frac{4}{\pi^2}\int_{0}^{\frac{1}{2}|k_{Fj}-\underline{k}|}\theta(k_{Fj}-\underline{k})dt\,t^2\left(2{}^V\Gamma_{ij}^{dir}
 -\frac{1}{2}\sum_m\xi_{mv}{}^{m}\Gamma_{ij}^{ex}\right)\right.\nonumber\\
  &&\left.-\frac{1}{2\pi^2}\int_{\frac{1}{2}|k_{Fj}-\underline{k}|}^{\frac{1}{2}(k_{Fj}+\underline{k})}dt\frac{t}{\underline{k}}(k_{Fj}^2-4t^2
  -\underline{k}^2+4t\underline{k})\left(2{}^V\Gamma_{ij}^{dir}-\frac{1}{2}\sum_m\xi_{mv}{}^{m}\Gamma_{ij}^{ex}\right)\right],\\
  &&\Sigma_i^V(\underline{k},E^{\ast}_i(k))\nonumber\\
  &=&\sum_{j=p,n}\theta(k_{Fj})\left[-\frac{4}{\pi^2}\int_{0}^{\frac{1}{2}|k_{Fj}-\underline{k}|}\theta(k_{Fj}-\underline{k})dt\frac{t^2}{E_{j\,av}^{\ast}(t)}\left(2{}^V\Gamma_{ij}^{dir} 
 -\frac{1}{2}\sum_m\xi_{mv}{}^{m}\Gamma_{ij}^{ex}\right)\right.\nonumber\\
  &&-\frac{1}{8\pi^2\underline{k}^3}\int_{\frac{1}{2}|k_{Fj}-\underline{k}|}^{\frac{1}{2}(k_{Fj}+\underline{k})}dt\frac{t}{E_{j\,av}^{\ast}(t)}(k_{Fj}^2-4t^2-\underline{k}^2 
 +4t\underline{k})\nonumber\\
  &&\left.\times(k_{Fj}^2+3\underline{k}^2-4t^2-4t\underline{k})\left(2{}^V\Gamma_{ij}^{dir}-\frac{1}{2}\sum_m\xi_{mv}{}^{m}\Gamma_{ij}^{ex}\right)\right],
\end{eqnarray}
and those in the complete pv representation read 
\begin{eqnarray}
	&&\Sigma_i^S(\underline{k},E^{\ast}_i(k))\nonumber\\
	&=&\sum_{j=p,n}\theta(k_{Fj})\left\{\frac{2}{\pi^2}\int_0^{\frac{1}{2}|k_{Fj}-\underline{k}|}\theta(k_{Fj}-\underline{k})dt\,t^2\frac{M_j^{\ast}}{E_{j~av}^{\ast}(t)}\left[\tilde{g}_{ij}^S 
 -\frac{2k^2}{(M_j^{\ast}+M_i^{\ast})^2}\left({}^{PV}g_{ij}^{ex}+{}^{\tilde{PV}}g_{ij}^{dir}\right)\right]\right.\nonumber\\
	&&+\frac{1}{4\pi^2}\int_{\frac{1}{2}|k_{Fj}-\underline{k}|}^{k_{Fj}+\underline{k}}dt\,\frac{t}{\underline{k}}\frac{M_j^{\ast}}{E_{j~av}^{\ast}(t)}(k_{Fj}^2-4t^2-\underline{k}^2 
 +4t\underline{k})\nonumber\\
	&&\left.\times\left[\tilde{g}^S_{ij}-\frac{1}{2(M_j^{\ast}+M_i^{\ast})^2}(k_{Fj}^2+3\underline{k}^2-4t^2-4t\underline{k})\left({}^{PV}g_{ij}^{ex}+{}^{\tilde{PV}}g_{ij}^{dir}\right)\right]\right\},\\
	&&\Sigma_i^0(\underline{k},E^{\ast}_i(k))\nonumber\\
	&=&\sum_{j=p,n}\theta(k_{Fj})\left\{\frac{2}{\pi^2}\int_0^{\frac{1}{2}|k_{Fj}-\underline{k}|}\theta(k_{Fj}-\underline{k})dt\,t^2\left[\tilde{g}_{ij}^0+\frac{2\underline{k}^2}{(M_j^{\ast} 
  +M_i^{\ast})^2}\frac{E_i^{\ast}(k)}{E_{j~av}^{\ast}(t)}\left({}^{PV}g_{ij}^{ex}+{}^{\tilde{PV}}g_{ij}^{dir}\right)\right]\right.\nonumber\\
	&&+\frac{1}{4\pi^2}\int_{\frac{1}{2}|k_{Fj}-\underline{k}|}^{\frac{1}{2}(k_{Fj}+\underline{k})}dt\,\frac{t}{\underline{k}}(k_{Fj}^2-4t^2-\underline{k}^2+4t\underline{k})\nonumber\\
	&&\left.\times\left[\tilde{g}_{ij}^0+\frac{1}{2(M_j^{\ast}+M_i^{\ast})^2}\frac{E_i^{\ast}(k)}{E_{j~av}^{\ast}(t)}(k_{Fj}^2+3\underline{k}^2-4t^2 
  -4t\underline{k})\left({}^{PV}g_{ij}^{ex}+{}^{\tilde{PV}}g_{ij}^{dir}\right)\right]\right\},\\
	&&\Sigma_i^V(\underline{k},E^{\ast}_i(k))\nonumber\\
	&=&\sum_{j=p,n}\theta(k_{Fj})\left\{\frac{2}{\pi^2}\int_0^{\frac{1}{2}|k_{Fj}-\underline{k}|}\theta(k_{Fj}-\underline{k})dt\,t^2\frac{1}{E_{j~av}^{\ast}(t)}\right.\nonumber\\
	&&\times\left[\frac{2}{(M_j^{\ast}+M_i^{\ast})^2}(M_j^{\ast2}-E_{j~av}^{\ast}(t)E_i^{\ast}(k))\left({}^{PV}g_{ij}^{ex}+{}^{\tilde{PV}}g_{ij}^{dir}\right)+\tilde{g}_{ij}^V\right]\nonumber\\
	&&+\frac{1}{4\pi^2}\int_{\frac{1}{2}|k_{Fj}-\underline{k}|}^{\frac{1}{2}(k_{Fj}+\underline{k})}dt\,\frac{t}{\underline{k}}\frac{1}{E_{j~av}^{\ast}(t)}(k_{Fj}^2-4t^2 
  -\underline{k}^2+4t\underline{k})\nonumber\\
	&&\times\left[\frac{2}{(M_j^{\ast}+M_i^{\ast})^2}(M_j^{\ast2}-E_{j~av}^{\ast}(t)E_i^{\ast}(k))\left({}^{PV}g_{ij}^{ex}+{}^{\tilde{PV}}g_{ij}^{dir}\right)\right.\nonumber\\
	&&\left.\left.+\frac{1}{4\underline{k}^2}(k_{Fj}^2+3\underline{k}^2-4t^2-4t\underline{k})\tilde{g}_{ij}^V\right]\right\},
\end{eqnarray}
where 
\begin{eqnarray}
	\tilde{g}_{ij}^S&\equiv&4\left({}^Sg_{ij}^{dir}+{}^{\tilde{S}}g_{ij}^{ex}\right)-\left({}^Sg_{ij}^{ex}+{}^{\tilde{S}}g_{ij}^{dir}\right)+4\left({}^Ag_{ij}^{ex}+{}^{A}g_{ij}^{dir}\right)\nonumber\\
	&&-\frac{1}{(M_i^{\ast}+M_j^{\ast})^2}\left\{M_i^{\ast2}+M_j^{\ast2}-2E_i^{\ast}(k)E_{j~av}^{\ast}(t)\right\}\left({}^{PV}g_{ij}^{ex}+{}^{\tilde{PV}}g_{ij}^{dir}\right),\\
	\tilde{g}_{ij}^0&\equiv&\left({}^{S}g_{ij}^{ex}+{}^{\tilde{S}}g_{ij}^{dir}\right)-2\left({}^{A}g_{ij}^{ex}+{}^{A}g_{ij}^{dir}\right)\nonumber\\
	&&+\frac{1}{(M_i^{\ast}+M_j^{\ast})^2E_{j~av}^{\ast}(t)}\left\{2E_i^{\ast}(k)(M_j^{\ast2}-E_{j~av}^{\ast}(t)E_i^{\ast}(k))\right.\nonumber\\
	&&\left.-E_{j~av}^{\ast}(t)(M_j^{\ast2}-M_i^{\ast2})\right\}\left({}^{PV}g_{ij}^{ex}+{}^{\tilde{PV}}g_{ij}^{dir}\right),\\
	\tilde{g}_{ij}^V&\equiv&\left({}^{S}g_{ij}^{ex}+{}^{\tilde{S}}g_{ij}^{dir}\right)-2\left({}^{A}g_{ij}^{ex}+{}^{A}g_{ij}^{dir}\right)+\frac{2k^2+M_i^{\ast2}-M_j^{\ast2}}{(M_j^{\ast} 
  +M_i^{\ast})^2}\left({}^{PV}g_{ij}^{ex}+{}^{\tilde{PV}}g_{ij}^{dir}\right).
\end{eqnarray}

%%%%%%%%%%%%%%%%%%%%%%%%%%%%%%%%%%%%%%%%%%%%%%%%%
%
\section{Potential energy\label{sec:Potential_energy}}

The potential energy, $\left<\hat{\cal V}_{ij}\right>$, in Eq. (\ref{eq:Vij}) can be rewritten as
\begin{eqnarray}
	\left<\hat{\cal V}_{ij}\right>&=&\frac{1}{(2\pi)^4}\int p^2dp\int P^2dP\int_{-1}^1 d\mu^{\prime}\frac{\tilde{\Gamma}_{ij}}{2E_j^{\ast}(\frac{1}{2}{\bm P}-{\bm p}) 
  E_i^{\ast}(\frac{1}{2}{\bm P}+{\bm p})} \nonumber \\
	&&\times\theta\left(k_{Fi}^2-\frac{1}{4}P^2-Pp\mu^{\prime}-p^2\right)\theta\left(k_{Fj}^2-\frac{1}{4}P^2+Pp\mu^{\prime}-p^2\right) , \label{eq:Vij3}
\end{eqnarray}
with $p\equiv|{\bm p}|$, $P\equiv|{\bm P}|$ and 
$\mu^{\prime} \equiv \hat{\bm P}\cdot\hat{\bm p}$. 
This involves the three-dimensional integral, and, at each mesh of ($p$, $P$, $\mu^{\prime}$), we need considerable calculation time for obtaining $\tilde{\Gamma}_{ij}$. 
We thus approximate the integral, using the angle-averaged approximation, as discussed in the Appendix \ref{sec:Self_energy_S0V}.  

First, using the angle-averaged, effective energy, $E^{\ast}_{i \, av}$, at fixed $p$ and $P$, which is defined as 
\begin{equation}
	E_{i \, av}^{\ast}\left(\frac{1}{2}{\bm P}\pm{\bm p}\right)\equiv\int d\mu^{\prime}\,E_{i}^{\ast}\left(\frac{1}{2}{\bm P}\pm{\bm p}\right)\left(\int d\mu^{\prime}\right)^{-1} \ ,
\end{equation}
the integral in Eq. (\ref{eq:Vij3}) can be reduced to the two-dimensional one with respect to the variables, $p$ and $P$.  
Next, by introducing an averaging of the total momentum squared of the interacting two nucleons in matter, $P_{av}^2$, at fixed $p$, which is calculated as 
\begin{equation}
	P_{av}^2\equiv \int P^4dP\int d\mu^{\prime}\left(\int P^2dP\int d\mu^{\prime}\right)^{-1} \ ,
\end{equation}
the two-dimensional integral can be reduced further to the one-dimensional one with respect to the variable, $p$.  

These approaches have for the first time been used in the DBHF calculation by Horowitz and Serot \cite{Serot,Horowitz} to 
study the properties of symmetric nuclear matter around $n_B^0$.  It is thus very important to examine if these approximations are applicable even to the 
DBHF calculation in isospin-asymmetric nuclear matter at higher density.

%\clearpage
%
%%%%%%%%%%%%% Bibliography %%%%%%%%%%%%%%%%%%%%%%%%%%%%%%%%%
%
%\newpage

%

%\newpage %Just because of unusual number of tables stacked at end

\end{document}